\documentclass[aps,prl,superscriptaddress,floatfix,showpacs] {revtex4}
\usepackage{amsmath,amssymb,epsfig,pstricks,bm,graphics,alltt}

\begin{document}

\title{Temperature induced in-gap states in the band structure and the insulator-metal transition in LaCoO$_3 $}

\author {S.G.~Ovchinnikov}
\email {sgo@iph.krasn.ru}
\affiliation {\mbox{Institute of Physics, Siberian Branch Russian Academy of Sciences, Krasnoyarsk 660036, Russia}}
\affiliation {\mbox{Siberian Federal University, Krasnoyarsk 660041, Russia}}
\affiliation {\mbox{Siberian Aerospace University, Krasnoyarsk 660014, Russia}}
\author {Yu.S.~Orlov}
\email {jso.krasn@mail.ru}
\affiliation {\mbox{Institute of Physics, Siberian Branch Russian Academy of Sciences, Krasnoyarsk 660036, Russia}}
\author {I.A.~Nekrasov}
\email {nekrasov@iep.uran.ru}
\affiliation {\mbox{Institute for Electrophysics, Russian Academy of Sciences-Ural Division, Yekaterinburg 620016, Russia}}
\author {Z.V.~Pchelkina}
\email {pzv@ifmlrs.imp.uran.ru}
\affiliation {\mbox{Institute of Metal Physics, Russian Academy of Sciences-Ural Division, Yekaterinburg 620990, Russia}}


\begin{abstract}
For many years a spin-state transition at $T \approx 100 K$ and insulator - metal transition (IMT) at $T_{IMT}  \approx 600 K$ in LaCoO$_3 $ remains a mystery. Small low-spin - high-spin spin gap $\Delta _S  = E\left( {HS} \right) - E\left( {LS} \right) \sim 100 K$ results in the spin-state transition. The large charge gap $2E_a  \approx 2300 K$ ($E_a $ is the activation energy) vs. $\Delta _S $ and $T_{IMT} $ implies that LaCoO$_3 $ is not a simple narrow-gap semiconductor. Here we explain both the spin-state and IMT on the same footing. We obtain strong temperature dependent band structure in LaCoO$_3 $ by the LDA+GTB method that incorporates strong electron correlations, covalence and spin-orbital interaction exactly inside the CoO$_6 $ cluster and the interclaster hopping between different multielectron configurations by perturbation theory for Hubbard X-operators.
\end{abstract}


\maketitle

The perovskite-oxide LaCoO$_3 $ has been studied intensely for many years due its unique magnetic properties and related insulator-metal transition (IMT) \cite{Jonker, Raccah Goodenough}. A gradual appearance of the paramagnetism above $50 K$ from the nonmagnetic ground state in called the spin-state transition. Goodenough was the first who suggested that instead of the Hund's rule dictated high spin (HS) $S = 2$ the strong crystalline electric field results in the low spin (LS) $S = 0$ state for $d^6 $ configuration of the Co$^{3 + } $ ion, and the energy difference is very small with the spin gap $\Delta _S  = E\left( {HS} \right) - E\left( {LS} \right) \sim 100 K$. The thermal population of the HS state provides the sharp increase of the magnetic susceptibility $\chi $ with a maximum around $100 K$. The nature of the excited spin state of Co$^{3 + } $ above the singlet $^1 A_{1g} $ has been under debate (see for a recent review \cite{Ivanova NB review}). Besides the original $^5 T_{2g} $ HS state with the $t_{2g}^4 e_g^2 $ configurations \cite{Raccah Goodenough} there were many indications on the intermediate spin (IS) $S = 1$ $^3 T_{1g} $ state. The two stage model has been proposed with the LS-IS transition at $100 K$ and IS-HS transition at $550-600 K$ \cite{two stage model 1, two stage model 2}. Recent electronic spin resonance (ESR) \cite{ESR Noguchi}, X-ray absorption spectroscopy (XAS) and X-ray magnetic circular dichroism (XMCD) \cite{XMCD Haverkort} experiments prove that the lowest excited state is really the HS. Nevertheless the $^5 T_{2g} $ term is splitted by the spin-orbital interaction in the low energy triplet with effective moment $\tilde J = 1$, and higher energy sublevels with $\tilde J = 2$ and $\tilde J = 3$ \cite{Ropka Radwanski}. The large difference between the spin excitation gap $\Delta _S $ and the charge gap given by the activation energy for electrical conductivity $E_a  \approx 0,1 eV$ at low $T$ indicates that $LaCoO_3 $ is not a simple band insulator \cite{is not a simple band insulator/exp resistance and electrical conductivity}. The second shallow maximum in $\chi $ near $500-600 K$ is often related to the insulator-metal transition. Surprisingly for the IMT electrical conductivity $\sigma $ does not seem to show any noticeable anomaly at this temperature \cite{is not a simple band insulator/exp resistance and electrical conductivity}. Moreover the discrepancy between the large charge gap $2E_a  \approx 2300 K$ and the insulator-metal transition temperature $T_{IMT}  \approx 600 K$ implies that the IMT cannot be simply argued in terms of a narrow-gap semiconductor \cite{cannot be simply argued in terms of a narrow-gap semiconductor/experimental value Eg}. Here, we solved this problem by calculating the electronic band structure in the regime of strong electron correlations. We consider electron as the linear combination of quasiparticles (QP) given by excitations between the different multielectron configurations obtained by exact diagonalization of the CoO$_6 $ cluster. With the Hubbard operators constructed within the exact cluster eigenstates we can calculate the QP band structure for the infinite lattice. The QP spectral weight is determined by the occupation numbers of the local multielectron configurations. We find that the thermal population of different sublevels of the ${}^5T_{2g} $ HS term splitted by the spin-orbital interaction results both in the spinstate transition and also in some new QP excitations. Of particular importance is the hole creation QP from the initial $d^6 $ HS into the $d^5 $ HS term, this QP appears to form the in-gap state inside the large charge-transfer gap $E_g  \approx 1,5 eV$. The intercluster hopping transforms this local QP into the in-gap band that lies just under the bottom of empty conductivity band and provides the insulating gap $2E_a  \approx 0,2 eV$ at $T = 100 K$. It bandwidth increases with $T$, and verlaps with the conductivity band at $T = T_{IMT}  = 587 K$ resulting in the IMT. Hence our approach allows to treat both the low $T$ spin-state transition and the high $T$ IMT on the same footing.

LaCoO$_3 $ as well as other strongly correlated oxides is a difficult problem for the ab initio band theory. The LDA calculations \cite{LDA calculations} incorrectly predict a metal for paramagnetic LaCoO$_3 $. Various methods have been applied to study effect of correlations on the LaCoO$_3 $ electronic structure: LDA+U or GGA+U \cite{LDA+U,Various methods1,Various methods2,Various methods3}, dynamical mean-field theory \cite{DMFT Muller-Hartmann}. Recent variational cluster approximation (VCA) calculation \cite{VCA Eder} based on the exact diagonalization of the CoO$_6 $ cluster gives a reasonably accurate description of the low temperature properties: the insulating nature of the material, the photoelectron spectra, the LS-HS spin-state transition. The main deficiency of the VCA is the failure to reproduce the high temperature anomalies in the magnetic and electronic properties associated with the IMT. The exact diagonalization of the multielectron Hamiltonian for a finite cluster provides a reliable general overview of the electronic structure of the correlated materials \cite{dop1}. To incorporate the lattice effect several versions of the cluster perturbation theory are known \cite{dop2,dop3}. To calculate the band dispersion in the strongly correlated material one has to go beyond the local multielectron language. The natural tool to solve this problem is given by the Hubbard  X-operators $X_f^{pq}  = \left| p \right\rangle \left\langle q \right|$ constructed with the CoO$_6 $ cluster eigenvectors $\left| p \right\rangle $ at site $\vec R_f $. All effects of the strong Coulomb interaction, spin-orbit coupling, covalence and the crystal field inside the CoO$_6 $ cluster are included in the set of the local eigenstates $E_p $. Here $p$ denotes the following quantum numbers: the number of electrons (both 3d Co and p of O), spin $S$ and pseudoorbital moment $\tilde l$ (or the total pseudomoment $\tilde J$ due to spin-orbit coupling), the irreducible representation in the crystal field. A relevant number of electrons is determined from the electroneutrality, for stoichiometric LaCoO$_3 $ $n = 6$. In the pure ionic model the corresponding energy level scheme for $d^6 $ $Co^{3 + } $ has been obtained in \cite{Ropka Radwanski}. Due to the covalence there is admixture of the ligand hole configurations $d^{n + 1} \underline L $ and $d^{n + 2} \underline L ^2 $ that is very well known in the X-ray spectroscopy \cite{ligand hole configurations well known in the X-ray spectroscopy}. Contrary to spectroscopy the electronic structure calculations require the electron addition and removal excitations. For LaCoO$_3 $ it means the $d^5 $ and $d^7 $ configurations. The total low energy Hilbert space is shown in the fig.~1. Here the energy level notations are the same as in the ionic model \cite{Ropka Radwanski} but all eigenstate contains the oxygen hole admixture due to the covalence effect. The calculation of the $n = 5,\;6,\;7$ eigenvectors for CoO$_6 $ cluster with the spin-orbit coupling and the Coulomb interaction has been done in \cite{Orlov Ovchinnikov}.

Following the generalized tight-binding (GTB) method \cite{GTB} and its ab initio version LDA+GTB \cite{LDA+GTB} we consider the electron annihilation operator at site $f$, orbital state $\lambda $ (Co d or O p) and spin projection $\sigma $ as the linear combination of the Hubbard operators
\begin{equation}
\label{annihilation operator as the linear combination of the Hubbard operators}
a_{f\lambda \sigma }  = \sum\limits_n {\gamma _{\lambda \sigma } (n)} X_f^n
\mbox{.}
\end{equation}
Here $n$ numerates different pairs of indexes in $X^{pq}  = \left| p \right\rangle \left\langle q \right|$ to describe the excitation from the initial state $\left| q \right\rangle $ to the final one $\left| p \right\rangle $. The matrix element $\gamma _{\lambda \sigma } (n)$ is calculated straightforwardly as the eigenvectors $\left\{ {\left| p \right\rangle } \right\}$ are known from the exact diagonalization of CoO$_6 $ cluster. In the X-operator representation the initial multielectron and multiorbital Hamiltonian has the same operator structure as the Hubbard model
\begin{equation}
H = \sum\limits_{fp} {E_p X_f^{pp} }  + \sum\limits_{fgnm} {t_{fg}^{mn} } \mathop {X_f^m X_g^n }\limits^ + 
\mbox{.}
\end{equation}
Here $E_p $ is the local eigenstate. All intercluster Coulomb interactions and hopping are included in the $\hat t_{fg} $ matrix and are treated perturbatively. The electron Green function 
\begin{equation}
G_{\lambda \lambda '\sigma } (k,\omega ) = \left\langle {\left\langle {{a_{k\lambda \sigma } }}
 \mathrel{\left | {\vphantom {{a_{k\lambda \sigma } } {a_{k\lambda '\sigma }^ +  }}}
 \right. \kern-\nulldelimiterspace}
 {{a_{k\lambda '\sigma }^ +  }} \right\rangle } \right\rangle _\omega   = \sum\limits_{nm} {\gamma _{\lambda \sigma } (n)} \gamma _{\lambda '\sigma } (m)D^{mn} (k,\omega )
\mbox{, }
 D_{fg}^{mn}  = \left\langle {\left\langle {{X_f^m }}
 \mathrel{\left | {\vphantom {{X_f^m } {\mathop {X_g^n }\limits^ +  }}}
 \right. \kern-\nulldelimiterspace}
 {{\mathop {X_g^n }\limits^ +  }} \right\rangle } \right\rangle
\mbox{.}
\end{equation}
Its local part is found exactly
\begin{equation}
G_{\lambda \sigma }^{(0)} (k,\omega ) = \sum\limits_n {\left| {\gamma _{\lambda \sigma } (n)} \right|} ^2 \frac{{F(n)}}{{\omega  - \Omega _n }}
\mbox{.}
\end{equation}
Here $\Omega _n  = E_p \left( {N + 1} \right) - E_q \left( N \right)$ is the QP energy, $F(n) = \left\langle {X_f^{pp} } \right\rangle  + \left\langle {X_f^{qq} } \right\rangle $ is the filling factor. This factor provides zero spectral weight for the excitations between two unoccupied states. The non-zero contribution to the Green function requires a participation of at least one occupied state in the QP $n = (p,q)$.

The electron removal spectrum determines the top of the valence band, the corresponding QP are shown in the fig.~1 by thin solid lines as the excitation from the $^1 A_{1g} $ $d^6 $ singlet in the $^2 T_{2g} $ $d^5 $ states with $\tilde J = {1 \mathord{\left/
 {\vphantom {1 2}} \right.
 \kern-\nulldelimiterspace} 2}$ and $\tilde J = {3 \mathord{\left/
 {\vphantom {3 2}} \right.
 \kern-\nulldelimiterspace} 2}$. There energies are
\begin{equation}
\Omega _{V1}  = E\left( {d^6 ,\;^1 A_1 } \right) - E\left( {d^5 ,\;^2 T_2 ,\;\tilde J = {1 \mathord{\left/
 {\vphantom {1 2}} \right.
 \kern-\nulldelimiterspace} 2}} \right)
\mbox{, }
\Omega _{V2}  = E\left( {d^6 ,\;^1 A_1 } \right) - E\left( {d^5 ,\;^2 T_2 ,\;\tilde J = {3 \mathord{\left/
 {\vphantom {3 2}} \right.
 \kern-\nulldelimiterspace} 2}} \right)
\mbox{.}
\end{equation}
The bottom of empty conductivity band has the energy
\begin{equation}
\Omega _C  = E\left( {d^7 ,\;^2 E} \right) - E\left( {d^6 ,\;^1 A_1 } \right)
\mbox{.}
\end{equation}
All these bands have nonzero QP spectral weight. The intercluster hopping results in the dispersion, $\Omega _n  \to \Omega _n (k)$. In a simplest Hubbard-1 approximation for the $\hat t_{fg} $ Green function is given by
\begin{equation}
\hat D^{ - 1} (k,\omega ) = \hat D_0^{ - 1} (\omega ) - \hat t(k)
\mbox{,}
\end{equation}
the band structure is obtained from the dispersion equation
\begin{equation}
\label{dispersion equation}
\det \left\| {{{\delta mn\left( {E - \Omega _n } \right)} \mathord{\left/
 {\vphantom {{\delta mn\left( {E - \Omega _n } \right)} {F(n) - t^{mn} (k)}}} \right.
 \kern-\nulldelimiterspace} {F(n) - t^{mn} (k)}}} \right\| = 0
\mbox{.}
\end{equation}
In the LDA+GTB version the single-site electron energies as well as the interatomic hopping integrals of the p-d model has been obtained by Wannier function projection technique \cite{Wannier function projection technique}. The LDA band structure of LaCoO$_3$ was obtained within linearized muffin-tin orbital (LMTO) basis set \cite{LMTO} using structural data for $T=5 K$ \cite{structural data}. The calculated bands are in good agreement with the previous results \cite{LDA calculations}. The bands resulted from the projection procedure for five Co d-orbitals and three O p-orbitals completely reproduce the initial LDA bands. The local coordinate system with the axis directing towards oxygen atoms forming the octahedron around Co atom was used. The intra-atomic Hubbard and Hund parameters $U$ and $J$ have been calculated within the constrained LDA method \cite{constrained LDA method}. The QP LDA+GTB band structure corresponds to the charge-transfer insulator \cite{The QP LDA+GTB band structure corresponds to the charge-transfer insulator} with the gap $E_g  \approx 1.5 eV$ (fig.~2) at $T = 0 K$. This gap value is rather close to the VCA gap \cite{VCA Eder} and the experimental value $E_g  \approx 1 eV$ \cite{cannot be simply argued in terms of a narrow-gap semiconductor/experimental value Eg}.

At finite temperature the thermal excitation over the spin-gap $\Delta _S $ into the $\tilde J = 1$ and over the gap $\Delta _S  + 2\tilde \lambda $ into the $\tilde J = 2$ sublevels of the HS ${}^5T_{2g} $ state occurs. We take $\Delta _S  = 140 K$ and $\tilde \lambda  = 185 K$ following6. Partial occupation of the excited HS states results in the drastically change of the QP spectrum. For $T = 0 K$ excitations from the $^1 A_1 $ $d^6 $ singlet in the lowest $^6 A_1 $ $d^5 $ term were forbidden due to spin conservation (the corresponding matrix element $\gamma _n  = 0$, see~(\ref{annihilation operator as the linear combination of the Hubbard operators})), and the excitation from $\left| {d^6 ,\;\tilde J = 1} \right\rangle $ in $\left| {d^5 ,\;^6 A_1 } \right\rangle $ has nonzero matrix element (shown by dashed line $\Omega _{V1}^ *  $ in the fig.~1) but zero filling factor as the excitation between two empty states. For $T \ne 0$ the filling factor for the $\Omega _{V1}^ *  $ and $\Omega _{V2}^ *  $ QP is non zero and is equal to the occupation number $n_1 $ and $n_2 $ of the states $\left| {d^6 ,\;\tilde J = 1} \right\rangle $ and $\left| {d^6 ,\;\tilde J = 2} \right\rangle $ correspondingly. The energies of these QP are
\begin{equation}
\Omega _{V1}^ *   = E\left( {d^6 ,\;^5 T_{2g} ,\;\tilde J = 1} \right) - E\left( {d^5 ,\;^6 A_1 } \right)
\mbox{, }
\Omega _{V2}^ *   = E\left( {d^6 ,\;^5 T_{2g} ,\;\tilde J = 2} \right) - E\left( {d^5 ,\;^6 A_1 } \right)
\mbox{.}
\end{equation}
The energies of these QP appear to be slightly below the bottom of the conductivity band, see DOS at finite temperature in the fig.~2. Thus we have obtained that these temperature-induced QP states lies inside the charge-transfer gap, they are the in-gap states. Similar in-gap states are known to result from doping in the high temperature superconductors. The LaCoO$_3 $ is unique because the in-gap states are induced by heating. The chemical potential lies in the narrow gap $2E_a  \approx 0.2 eV$ at $T = 100 K$ between the in-gap states and conductivity band.

From the GTB dispersion equation~(\ref{dispersion equation}) it is clear that the in-gap bandwidth is proportional to the occupation numbers $n_1 $ and $n_2 $ of the excited HS states. With further temperature increase the in-gap bands $\Omega _{V1}^ *  $ and $\Omega _{V2}^ *  $ become wider and finally overlap with the conductivity band $\Omega _C $ (fig.~2) at $T = T_{IMT}  = 587 K$. It should be clarified that the IMT in LaCoO$_3 $ is not the thermodynamic phase transition, there is no any order parameter associated with the gap contrary to the classical IMT in VO$_2 $, $NiS$ etc.

Assuming that the carrier mobility is a smooth function of temperature we have estimated the temperature dependent conductivity as
\begin{equation}
\sigma (T) = \sigma _0 \exp \left( { - {{E_a (T)} \mathord{\left/
 {\vphantom {{E_a (T)} {kT}}} \right.
 \kern-\nulldelimiterspace} {kT}}} \right)
\end{equation}
with the temperature dependent $E_a (T)$ given by the DOS calculations. The fitting parameter $\sigma _0 $ was taken as $\sigma _0  = \sigma _{\exp } \left( {800 K} \right)$. The overall agreement of the calculated and experimental data from \cite{is not a simple band insulator/exp resistance and electrical conductivity} is clear (see fig.~3). The deviations at the $300-350 K$ we believe stem from some other interactions omitted in our calculations. There is similar anomaly in the thermal expansion coefficient in the same temperature range \cite{thermal expansion coefficient}, so may be the spin-phonon and electron-phonon interactions are responsible for this discrepancy.

We have also calculated the average moment $\vec J = \vec L + \vec S$ as $J_{av}  = \left\langle {\hat J^2 } \right\rangle ^{{\raise0.5ex\hbox{$\scriptstyle 1$}
\kern-0.1em/\kern-0.15em
\lower0.25ex\hbox{$\scriptstyle 2$}}} $ as function of temperature (fig.~4). Here
\begin{equation}
\left\langle {\hat J^2 } \right\rangle  = \sum\limits_n {\left\langle {n\left| {\hat J^2 } \right|n} \right\rangle } \left\langle {X^{nn} } \right\rangle
\mbox{.}
\end{equation}
The expected for HS value $J_{av}  \approx 2$ is reached only at $T \approx 1000 K$. In the region of the spin-state transition at $T \approx 100 K$ the value of $J_{av} $ is close to 1. May be it is the reason why so many experimentalists have obtained the IS $S = 1$ fitting their data by the state with definite spin. Temperature dependence of the magnetic susceptibility has the maximum at $T \approx 100 K$ similar to many previous works with LS-HS scenario \cite{is not a simple band insulator/exp resistance and electrical conductivity,Hoch}.

Thus, we find that a correct definition of the electron in strongly correlated system directly results in the in-gap states during the spin-state transition due to the thermal population of the excited HS states. Close to the spin-state temperature region the in-gap states determine the value of the activation energy $E_a  \approx 0.1 eV$. Further temperature increase results in larger in-gap bandwidth and smaller $E_a $, and finally $E_a  = 0$ at $T_{IMT}  = 587 K$. As concerns the weak maximum in the $\chi (T)$ close to the IMT it may be a small Pauli-type contribution from the itinerant carriers above $T_{IMT} $. We emphasize that instead of rather large difference in temperatures of the spin-state transition ($ \sim 100 K$) and the IMT ($600 K$) the underlying mechanism is the same and is induced by the thermal population of the excited HS states. \\

\textbf{Acknowledgements} \\

We acknowledge discussions with G.A. Sawatzky, M.W. Haverkort, S.V. Nikolaev and V.A. Gavrichkov. This work is supported by the Siberian Branch of RAS and Ural Branch of RAS integration project ¹40, Presidium RAS Program 5.7, RFBR grant ¹ 09-02-00171-a, RFBR grant ¹ 10-02-00251 and Dynasty Foundation.

{\sloppy

}
\newpage

\begin{figure}[ht]
\begin{center}
\includegraphics[width=.80\textwidth]{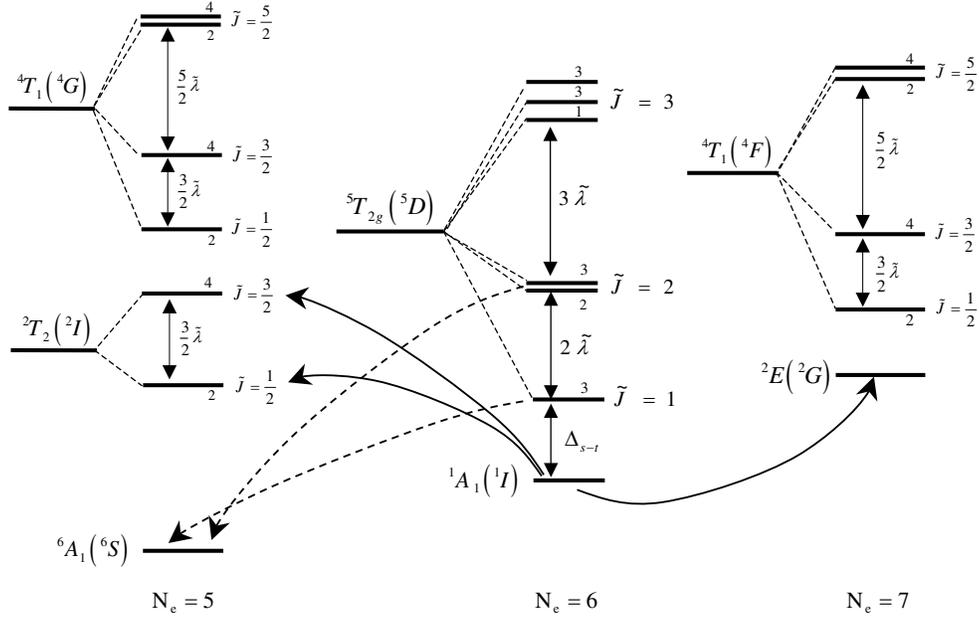}
\caption{The low energy part of the CoO$_6 $ cluster Hilbert state for the electron numbers $N_e  = 5,\;6,\;7$. Terms with given $N_e $ are the mixtures of $d^{N_e } $, $d^{N_e  + 1} \underline L $ and $d^{N_e  + 2} \underline L ^2 $ configurations. At $T = 0$ only the $N_e  = 6$ low spin term $^1 A_1 $ is occupied, the Fermi-type excitations from this term that form the top of the valence band ($d^6  \to d^5 $) and the bottom of the conductivity band ($d^6  \to d^7 $) are shown by the solid lines with arrow. The dashed lines denote the in-gap excitations with the spectral weight increasing with temperature due to the population the HS excited $d^6 $ terms.}
\end{center}
\end{figure}

{\sloppy

}
\newpage

\begin{figure}[ht]
\begin{center}
\includegraphics[width=.80\textwidth]{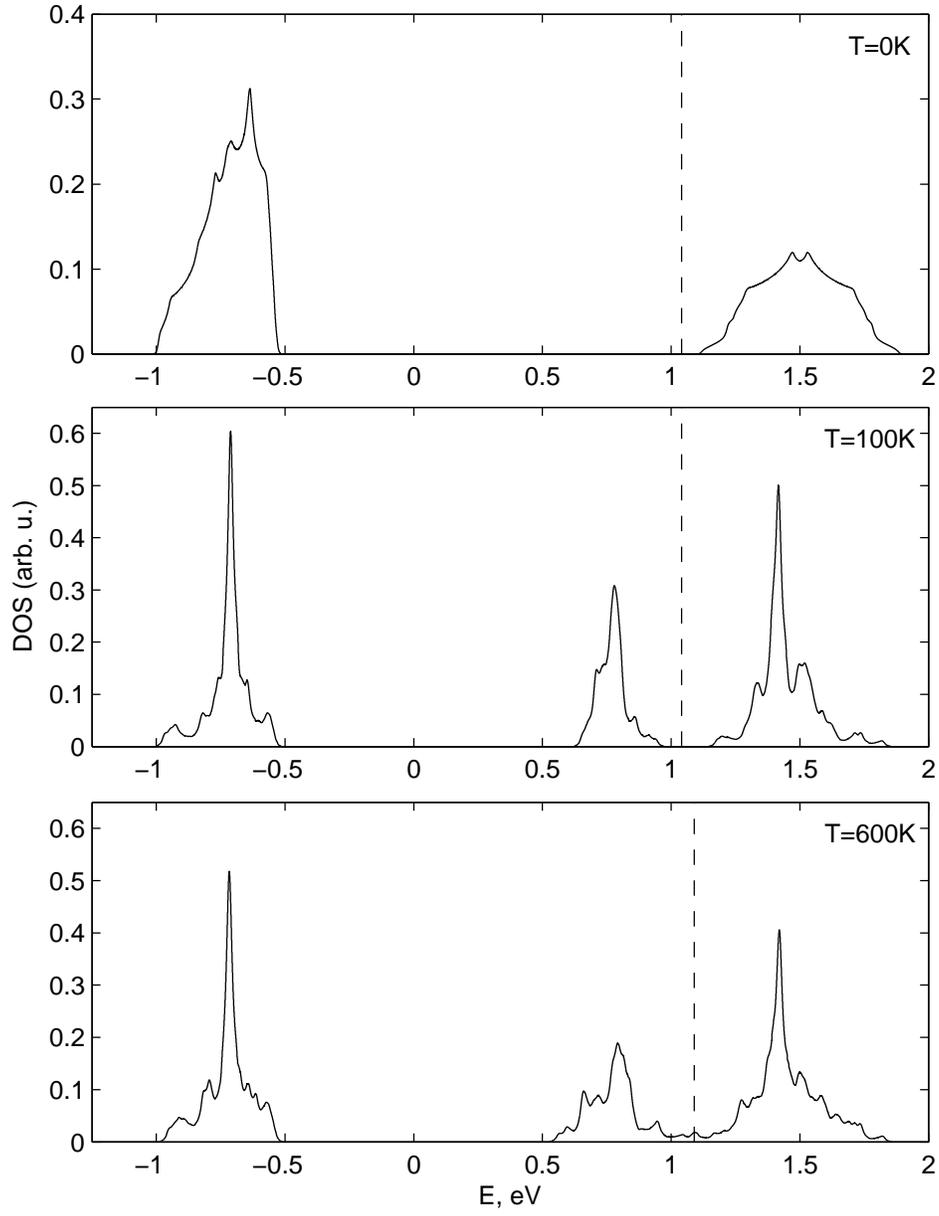}
\caption{Single particle density of states at different temperatures. At $T = 0 K$ LaCoO$_3 $ is the charge transfer insulator with the gap $E_g  \approx 1.5 eV$. At finite temperatures the in-gap band appears below the conductivity band with the temperature dependent activation energy. At $T = 100 K$ $E_a  \approx 0.1 eV$. At $T = T_{IMT}  = 587 K$ $E_a  = 0 eV$, and above $T_{IMT} $ the band structure is of the metal type.}
\end{center}
\end{figure}

{\sloppy

}
\newpage

\begin{figure}[ht]
\begin{center}
\includegraphics[width=.80\textwidth]{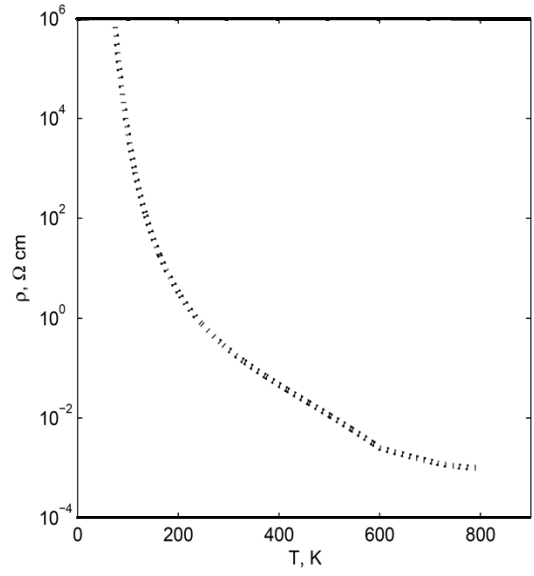}
\caption{The temperature dependence of the resistivity, solid line shows the experimental data \cite{is not a simple band insulator/exp resistance and electrical conductivity} and dashed line is calculated from Eq.(4).}
\end{center}
\end{figure}

{\sloppy

}
\newpage

\begin{figure}[ht]
\begin{center}
\includegraphics[width=.80\textwidth]{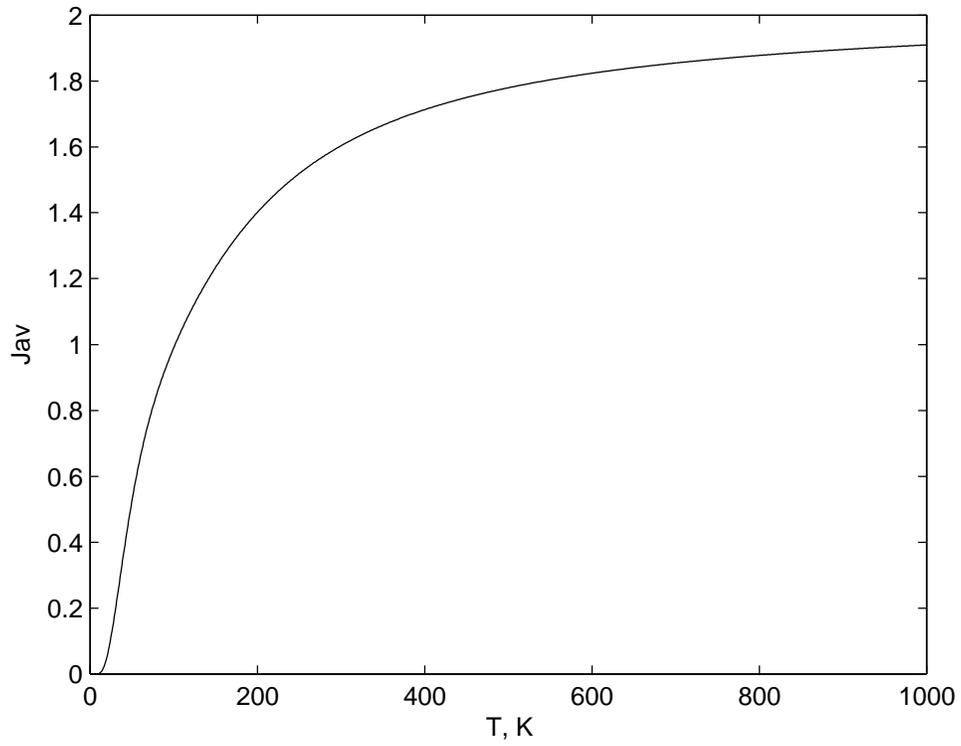}
\caption{The temperature dependence of the average local moment shows the spin-state transition from the nonmagnetic low spin state at $T = 0 K$ to the paramagnetic high spin state at finite temperature. Only at $T \approx 1000 K$ the moment is close to 2. In the region of the spin-state transition at $T \sim 100 K$ $J_{av}  \approx 1$.}
\end{center}
\end{figure}

\end{document}